# Charge Transport in DNA Segments with fractal structures

Huijie Yang[*], Fangcui Zhao, Chunchun Liu, Yingli Zhao, Wenxiu Yang and Beilai Hu

Institute of Physics, Nankai University, Tianjin 300071, China

## Abstract

By means of the concept of factorial moment the charge transfer rates in DNA segments with fractal structures are investigated. An analytical form for the electron transfer rate is obtained.

**PACS** numbers: 87.15-v, 87.14.Gg

---

[*] Correponding author, E-mail: huijieyangn@eyou.com

## I. Introduction

Electron transport, as a ground to a wide range of important biological processes in DNA, attracts special attentions for its fundamental physical interest and potential applications in DNA-based molecular technologies [1-2]. Two kinds of technologies are used to obtain the electron transport information in DNA. Direct or indirect electrical conductivity measurements on micrometer-long DNA ropes give an ambiguous result, the conductivity $\sigma$ ranging from $10^4 \Omega^{-1} cm^{-1}$ to $10^{-1} \Omega^{-1} cm^{-1}$ [3-7]. A much more reliable measurement technology based upon fluorescence quenching can trap a migrating electron at the acceptor site and monitor the charge transport by the yield of a chemical reaction accompanying this trapping process. Assuming the transfer rate can be characterized by an exponential law, $\exp(-\beta x)$, where $x$ is the donor-acceptor separation, the values of $\beta$ range from $0.1 \overset{\circ}{A}^{-1}$ to $1.4 \overset{\circ}{A}^{-1}$ [8-24]. To explain the contradictory results in a unified theoretical scheme two kinds of mechanisms are suggested in literature, i.e., coherent super-exchange and hoping process. If the energy levels of base pair stacking are much higher than that of the donor and the acceptor, coherent transport occurs, called coherent super-exchange, and the transfer rate decreases exponentially. Contrarily, the energy levels of base pair stacking are lower than or similar with that of the donor, hopping process occurs and the transfer rate decreases very slowly with the distance between the donor and the acceptor. Accordingly, the electron transport process is determined by the state of the DNA segment, which is related with the acceptor, the donor, the ratio of components, the DNA sequence and temperature, etc. [16,21,25-28]. As inherent factors, how the ratio of components and the sequence pattern of a DNA segment determine its conductivity is an essential problem to be investigated quantitatively in detail.

In this paper we introduce a method to describe quantitatively the walking fractal dimension of a short DNA segment ($\sim 10^2 bp$), by means of which we try to derive a quantitative relation between the complexities of a DNA segment with fractal structure and its electron transfer rate. This relation may be helpful for us to understand deeply the microscopic mechanisms for electron transport in DNA. It may also be useful for us to design a DNA segment with a specified value of electron transfer rate.

## II. Anomalous transport of an electron along a DNA segment

Two basic physical effects may contribute to the electron transport process. One is the diffusion of an electron along the DNA segment; the other is the quantum tunneling [23,29]. Investigations in literatures point out that the diffusion leads to the charge transport probability independent of the donor-acceptor distance, while the quantum tunneling leads to the exponential dependence of the probability on x. That is, the diffusion dominates the process of long distance electron transport. Considering a DNA segment with fractal structure, the electron transport can be treated as a diffusion process along one-dimensional fractal media. For one-dimensional media the value of geometric fractal dimension is $d_f = 1$. The couplings between the electrons and fractons induce the transport process. This transport process can be described completely by the anomalous transport equation, which reads [30],

$$\frac{\partial^{\gamma'} p(r,t)}{\partial t^{\gamma'}} = -A' r^{-\theta'} \frac{\partial p(r,t)}{\partial r} \quad (1)$$

Where $p(r,t)$ is the probability of finding an electron at position $r$ when time is $t$, assuming this electron is at $r = 0$ when time is zero. The two parameters $r', \theta'$ obey a formula as $\frac{\gamma'}{1+\theta'} = \frac{1}{d_w}$, in which the parameter $d_w$ is the value of walking fractal dimension. $A'$ is a positive value. The solution of this equation can be written as,

$$p(r,t) \propto t^{-d_s/2} \exp\left[-\left(a \cdot \frac{r}{t^{1/d_w}}\right)^{\alpha}\right]. \quad (2)$$

Where $a$ is a constant relating with component ratio of different kinds of base pairs, and $\alpha = \frac{d_w(1+\theta')}{d_w - (1+\theta')}$. The parameter $d_s = \frac{2d_f}{d_w}$ is the value of fracton spectrum dimension, describing the elementary vibration excitations in DNA. $\theta'$ is a parameter depending on the type of fractal media. The diffusion of an electron along a DNA segment with fractal structure can be regarded as a percolation process. The value of $\theta'$ should be zero for this kind of percolation fractal media. The corresponding diffusion constant is $D = \frac{<r(t)^2>}{t} \propto t^{\frac{2}{d_w}-1}$. Hence we can obtain the relation between the conductivity $\sigma$ and the walking fractal dimension $d_w$ as,

$$\sigma = \frac{e^2 n}{k_B T} D \propto \frac{e^2 n}{k_B T} \cdot t^{\frac{2}{d_w}-1}. \tag{3}$$

Where $e, n, k_B, T$ are the electricity quantity of an electron, the density of current carrier, Boltzmann constant and temperature, respectively.

### III. Determine the walking dimension with factorial moment

The DNA segments with potential electron transport applications are usually several hundreds base pair longs. How to derive the values of the walking fractal dimensions from such short a DNA segment is an essential problem to be solved at present time. In our recent papers, the concepts of factorial moment and delay register vector are used to describe the complexity of a short data record [31-34]. This method can describe successfully the long-range correlations embedded in short stride time series. And this method can also find coding regions in a DNA sequence based upon the correlation differences between the coding and non-coding regions. To describe the fractal structure characteristic of a short DNA segment, the procedure can be illustrated as below,

(1) $d$ successive nucleotides are regarded as a case. For a DNA segment with length $N$ we can construct totally $N-d+1$ cases.

(2) For *m*'th case we can reckon the number of occurrences of the nucleotides A and T (or C and G), denoted with $n_m$.

Consider a special condition where the energy levels of donor and/or acceptor are similar with that of A (or T) base pair stacking. Experiments show that an electron can pass through A and T nucleotides easily, while C and G hardly. As a roughness description of this diffusion difference we can employ the number of A and T as the diffusion distance in each case. And the length of a case can be regarded as diffusion time. That is, an electron can step forward to the next position in a time unit if the base pair is A-T or T-A, while it can only stay at the original position to induce chemical reactions in a time unit if the base pair is C-G or G-C. The average of the diffusion distance square is then,

$$R(d) = \langle n_m^2 \rangle = \frac{\sum_{m=1}^{N-d+1} n_m^2}{N-d+1} \tag{4}$$

For a diffusion process in a fractal media we have $R(d) \propto d^{2/d_w}$, according to which we can obtain the value of walking fractal dimension $d_w$. But for a short segment, the number of cases is not enough to guarantee a reliable value of $d_w$ from the viewpoint of statistical theory. The small number of cases will induce large statistical fluctuations. To dismiss statistical fluctuations due to finite cases effectively, we can investigate the corresponding 2-order factorial moments instead of the average of diffusion distance square i.e.

$$F_2(d) = \langle n_m(n_m-1) \rangle = \frac{\sum_{m=1}^{N-d+1} n_m(n_m-1)}{N-d+1} \tag{5}$$

Factorial moment theory proves that the relation of the 2-order factorial moments versus the length of a case can give us a reliable value of walking fractal dimension [35-37].

A very recent paper [38] gives a simple description of electron transport, where the electron has a backward and a forward transport probabilities denoted with $k_-$ and $k_+$ respectively. The results for $k_-/k_+ \sim 0$ can fit with experiments very well. It is consistent with this direct transport viewpoint presented here in a certain degree.

### IV. Electron transfer rate

For a DNA segment with homogenously positioned A and T, the value of walking dimension is $d_w = 1$. The formula (2) reduces to a Gaussian distribution function, which reads,

$$p(r,t) \propto t^{-1} \exp\left[-\left(a \cdot \frac{r}{t^{1/2}}\right)^{-\gamma}\right] \gamma \to \infty \quad (6)$$

On the contrary, for a DNA segment with deterministic positioned A and T (A and T aggregate together), the value of walking dimension is then $d_w = \infty$. The formula (2) reduces to an exponentially decay function as,

$$p(r,t) \propto \exp[-ar] \quad (7)$$

Generally, the electron transport process in a DNA segment with fractal structure behaves in a state between the above two states. Hence, we can obtain the electron transfer rate theoretically, which reads,

$$k_{et}(R, d_w, a, t) = \frac{\int_R^\infty p(r,t)dr}{\int_0^\infty p(r,t)dr} \quad (8)$$

Substituting (2) into (8),

$$k_{et}(R, d_w, a, t) = \frac{\int_R^\infty t^{-1/d_w} \exp\left[-a\left(rt^{-1/d_w}\right)^\alpha\right]dr}{\int_0^\infty t^{-1/d_w} \exp\left[-a\left(rt^{-1/d_w}\right)^\alpha\right]dr} \quad (9)$$

For two special conditions, we have,

$$k_{et}(R, d_w, a, t)\big|_{d_w=1} = 1 \left(R \leq \frac{t}{a}\right)$$

$$k_{et}(R, d_w, a, t)\big|_{d_w=\infty} = \exp(-aR) \quad (10)$$

The formula (10) is just the form of transfer rate employed to fit with experimental data for protein and DNA molecules in most literatures [39].

Therefore, there are two inherent factors determine the conductivity characteristic of a DNA segment. One is the sequence pattern described with the walking fractal dimension $d_w$. The other is the component ratio of base pairs A-T and T-A described with parameter $a$. With the increasing of the component ratio of base pairs A-T and T-A, the value of parameter $a$ decreases, and electron transport can reach longer and longer a distance.

From the experimental results we can estimate the ranges of the parameter values. Choosing the unit of length **nm**, the corresponding values of $a$ range from 1.0 to 14.0 for results in literatures. The time scale of an electron transport dynamics is femtosecond (**fs**), and the time scale for an electron transfer process is $ps$. Hence, the unit of $a$ is $\left(\dfrac{fs^{1/d_w}}{nm}\right)^{\alpha}$.

In Fig.(1) the relation of $k_{et}(R, d_w, a, t)$ versus $d_w$ is presented. With the decreasing of the value of fractal walking dimension $d_w$, the electron transfer rate increases rapidly and the relaxation time for the corresponding electron transport process decreases quickly. Estimation value for the relaxation time is $\sim 100 ps$, which is consistent with the experimental results [14,23].

In Fig.(2) the relation of $k_{et}(R, d_w, a, t)$ versus $R$ is presented also. With the increasing of the distance between the donor and the acceptor $R$, the electron transfer rate decreases rapidly. Different values of $d_w$ and $a$ influence the decrease speed significantly. The cooperation of the three parameters $a, d_w$ and $R$ determines the decay characteristic completely.

## V. Conclusion

In summary, by means of the theory of diffusion in fractal media we investigate the electron transport processes in DNA segments with fractal structures. A good electron transport characteristic is found for a homogenous DNA segment with a specified value of walking fractal dimension, that is, a big electron transfer rate and a short relaxation time. The sensitivity of the electron transport characteristic to the fractal property of a DNA segment can be a good basis for our designing a DNA molecule with expected indexes. Fractal characteristic of a DNA segment may be essential for the long distance electron transfer. The component ratio of a DNA segment determines the value of parameter $a$, which is essential for the electron transfer distance. Parameters $(d_w, a)$ can be employed as indexes for a DNA segment's conductivity. It may be an interesting thing to investigate experimentally the conductivity characteristics of short DNA segments with fractal structures.

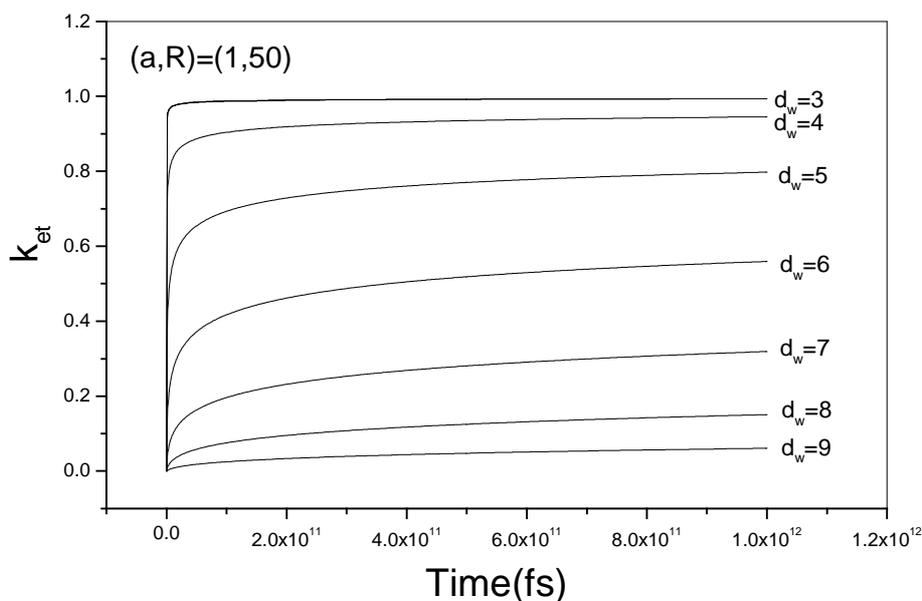

Fig.(1) Evolution of transfer rate for different fractal medium

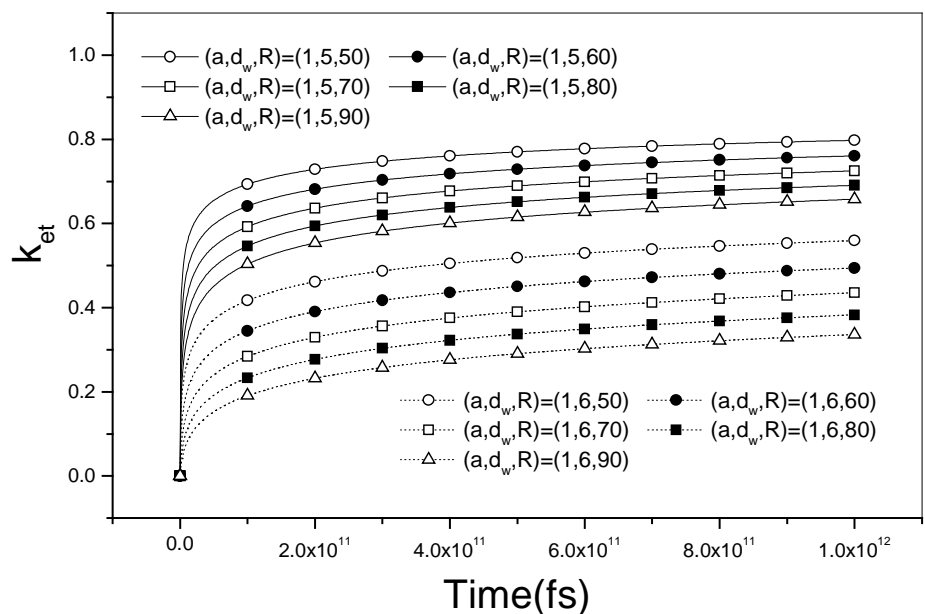

Fig.(2) Evolution of transfer rate for different transfer distance

## Acknowledgement

This work was supported partly by the National Science Foundation of China under Grant No.39770910. This work was also supported by the Doctor Promotion Foundation and the Post-doctor Foundation of Nankai University. One of the authors (H-J Yang) would like to thank professor Yizhong Zhuo, Xizhen Wu and Zhuxia Li for stimulating discussions and proposals.